\documentclass[a4paper,11pt]{article}
\usepackage{pos}
\usepackage{float}
\usepackage{subcaption}
\usepackage{lineno}

\title{Cosmic-ray energy reconstruction using machine learning techniques}
 \ShortTitle{CRs energy reconstruction using MLT}

\author[a]{A. Alvarado}
\author[b]{T. Capistr\'{a}n}
\author[c]{I. Torres}
\author[a]{J. R. Sacahu\'{i}}
\author*[d]{R. Alfaro}

\affiliation[a]{Instituto de Investigaci\'{o}n en Ciencias F\'{i}sicas y Matem\'{a}ticas USAC, Ciudad Universitaria, Zona 12, 01012, Guatemala}

\affiliation[b]{Instituto de Astronom\'{i}a, Universidad Nacional Aut\'{o}noma de M\'{e}xico, Ciudad de M\'{e}xico, Mexico}

\affiliation[c]{Instituto Nacional de Astrof\'{i}sica, \'{O}ptica y Electr\'{o}nica, Puebla, Mexico}

\affiliation[d]{Instituto de F\'{i}sica, Universidad Nacional Aut\'{o}noma de M\'{e}xico, Ciudad de M\'{e}xico, Mexico}

\onbehalf{for the HAWC Collaboration}

\emailAdd{daniel.alvarado004@gmail.com}
\emailAdd{tcapistran@astro.unam.mx}
\emailAdd{ibrahim@inaoep.mx}
\emailAdd{jrsacahui@profesor.usac.edu.gt}
\emailAdd{ruben@fisica.unam.mx}

\abstract{HAWC is a ground-based observatory consisting of $300$ water Cherenkov detectors, which observes the extensive air showers induced by cosmic rays from some TeV to a few PeV and, in particular, gamma rays from $300$~GeV to more than $100$~TeV. One of the crucial features required for a detector of extensive air showers is the estimation of the primary energy of the events to study the spectra of cosmic and gamma rays. For HAWC there are currently two gamma-ray energy estimators: one relies on a ground density parameter, while the other utilizes an artificial neural network. For the cosmic ray energy estimation, there is only one estimator based on maximum likelihood procedures and measurements of the lateral charge distribution of the events. It is worthwhile to update the cosmic-ray energy estimator due to recent improvements of the extensive air shower offline-reconstruction techniques in HAWC. Therefore, we implemented an artificial neural network to reconstruct the primary energy of hadronic events trained with several observables that characterize the air showers. We trained several models and evaluated their performance against the existing cosmic ray energy estimator. In this work, we present the features and performance of these models.}

\ConferenceLogo{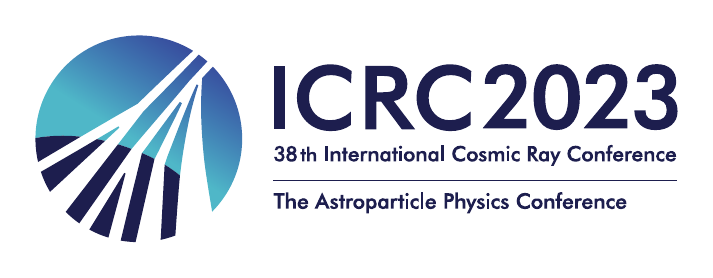}

\FullConference{%
38th International Cosmic Ray Conference (ICRC2023)\\
  26 July - 3 August, 2023\\
  Nagoya, Japan}

\begin{document}
\maketitle

\section{Introduction}
    \paragraph{}The High-Altitude Water Cherenkov (HAWC) gamma-ray observatory is located at an altitude of $4100$ m above sea level on the Sierra Negra volcano in Puebla, Mexico. HAWC consists of an array of $300$ water Cherenkov detectors, each instrumented with four photomultiplier tubes (PMTs), covering an area of $22000$~m$^2$~\citep{2023NIMPA105268253A}. HAWC operates nearly daily with a duty cycle exceeding 95\%, and it has a detection rate of approximately $25$~kHz. Given that the majority of the detected shower are cosmic ray, HAWC contributes significantly to the field. For instance, it can investigate the light mass group as reported in~\citep{2017PhRvD..96l2001A, PhysRevD.105.063021}, or the cosmic-ray anisotropy~\citep{anisotropyhawc}. One crucial factor in conducting such analyses is having an energy estimator for the primary particle. In previous works, it was used an energy estimator based on a maximum likelihood procedure and the measurements of the lateral charge distribution of the PMT with signals during the event (hereafter, we will refer to this estimation procedure as Likelihood)~\citep{2017PhRvD..96l2001A}. With recent improvements in the offline analysis of the HAWC software for shower reconstruction, the cosmic-ray energy estimator needs an update. To accomplish this task, in this contribution, we explore machine learning techniques, which will enable us to build a data-driven model. Section~\ref{sec:train} provides details of our model and its training, while, in section ~\ref{sec:test}, the performance  of the best trained model is presented and its is compared with the results for the likelihood technique. Finally, we summarize the results of this contribution and provide an overview of our future work (in section~\ref{sec:conclu}).

\section{Trained model sets}\label{sec:train}
    \paragraph{}For our analysis, we employed Monte Carlo simulation, which were computed using the standard procedure of HAWC. The CORSIKA package~\citep{corsika1998} (v740) was used to simulate the interaction between a primary particle and the atmosphere, as well as the resulting extensive air shower. FLUKA~\citep{Fluka2005} and QGSJet-II-04~\citep{qsjet} were employed as our low and high-energy hadronic interaction models, respectively. Eight specie\footnote{proton, helium, carbon, oxygen, neon, magnesium, silicon and iron} were simulated using a power-law energy spectrum, ranging from 5 GeV to 2 PeV, with an spectral index of $-2$. The GEANT4~\citep{GEANT4} package was utilized to simulate the interaction between secondary particles and the HAWC detector. Finally, the official HAWC's software was employed to reconstruct all shower event.

    \paragraph{}To train the neural networks, we utilized the TMVA package~\citep{TMVA} of ROOT~\citep{root}. The architecture of the neural network is defined as 14:10:10:1, where each number represents the number of artificial neurons in each layer. The first layer has fourteen neurons, which is the same number of input variables used. These variables contain information such as the lateral distribution charge (footprint) of the shower, its direction, the percentage of PMTs activated, and the distance of the shower core from the HAWC center. The second and third layers are the hidden layers with ten neurons each, and they use a sigmoid activation function. Finally, the last layer consists of one output neuron used to provide the prediction of the primary particle energy in units of $\log_{10}(E/$GeV). Among various model trained with different input variables, the selected fourteen variables yielded the best results. For training, we established 1000 epochs and utilized the Broyden-Fletcher-Goldfarb-Shannon (BFGS) learning model~\citep{TMVA}. During the training stage, two-thirds of the total proton events were used, while the remaining one-third of proton events were reserved for testing, approximately 5 millon of proton event were simulated. Protons were chosen for training the model because of their relative abundances in the intensity of cosmic rays, which is approximately 90\% of all detected particles~\cite{crsgaisser}.

    \paragraph{}Multiple sets of neural networks were trained, with each network operating on a different binning scheme. The binning scheme involves two parameters: the fractional signal of activated PMTs during the event ($f_{\text{hit}}$) and the position of the shower core relative to HAWC (for events with shower cores inside HAWC, we use the label in-HAWC, and for events at the border or outside HAWC, the label off-HAWC). Below is a description of three neural network sets used in this work:
    \begin{itemize}
        \item 1st set: A single trained network that predicts the energy for events, with shower core inside and outside of HAWC.
        \item 2nd set: Two trained networks are utilized. One network predicts the energy for in-HAWC events, while the other network estimates the energy for off-HAWC events.
        \item 3rd set: Three trained networks are employed for different $f_{\text{hit}}$ bins. The first network is designed for the low $f_{\text{hit}}$ bin range ($2.7\%$ - $22\%$), the second network for the medium $f_{\text{hit}}$ bin range ($22\%$ - $47\%$), and the third network for the high $f_{\text{hit}}$ bin range ($>47\%$).
    \end{itemize}

\begin{figure}
     \begin{subfigure}{0.49\textwidth}
         \includegraphics[width=\textwidth]{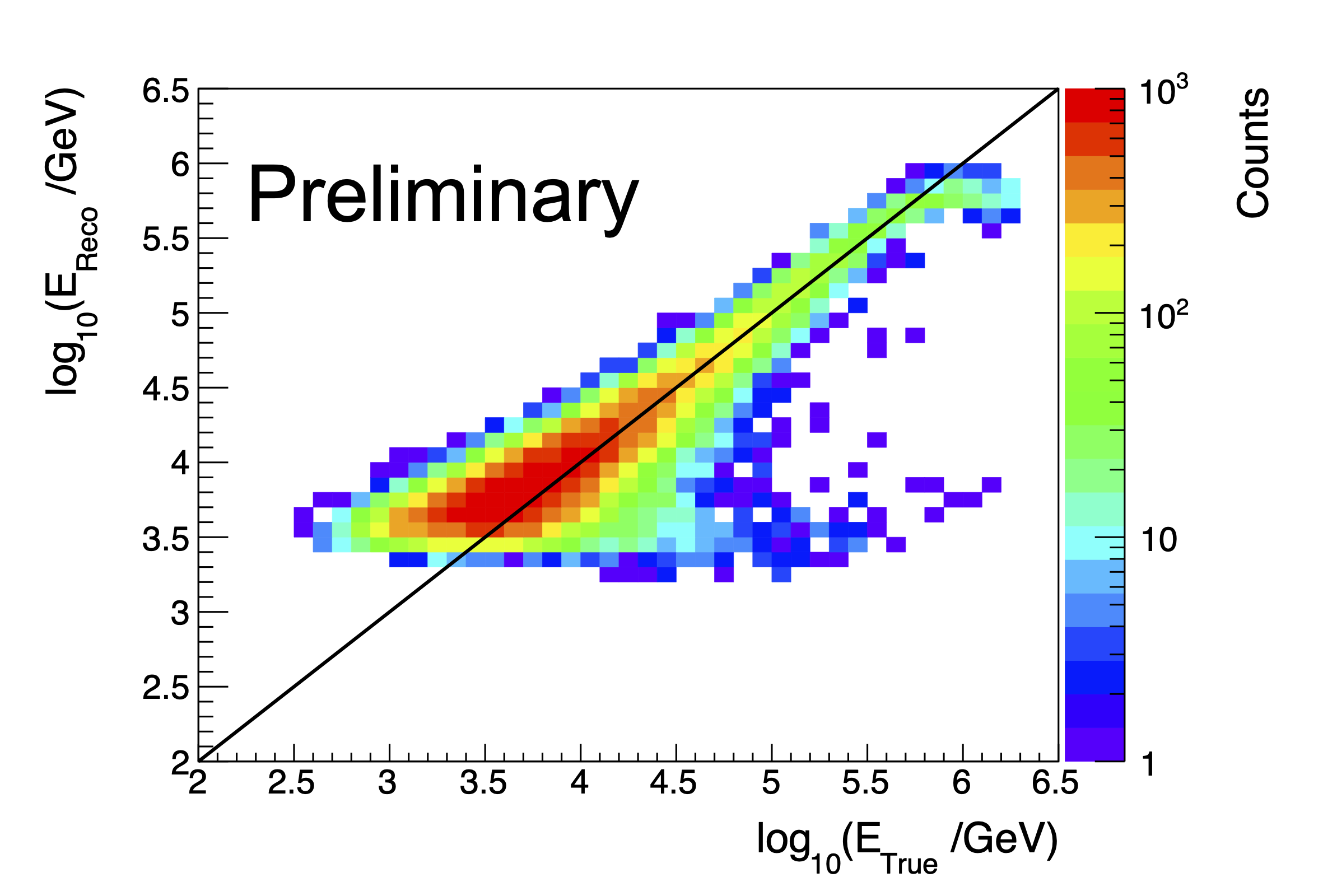}
         \caption{Likelihood}
         \label{fig:lhh2dsc}
     \end{subfigure}
     \hfill
     \begin{subfigure}{0.49\textwidth}
         \includegraphics[width=\textwidth]{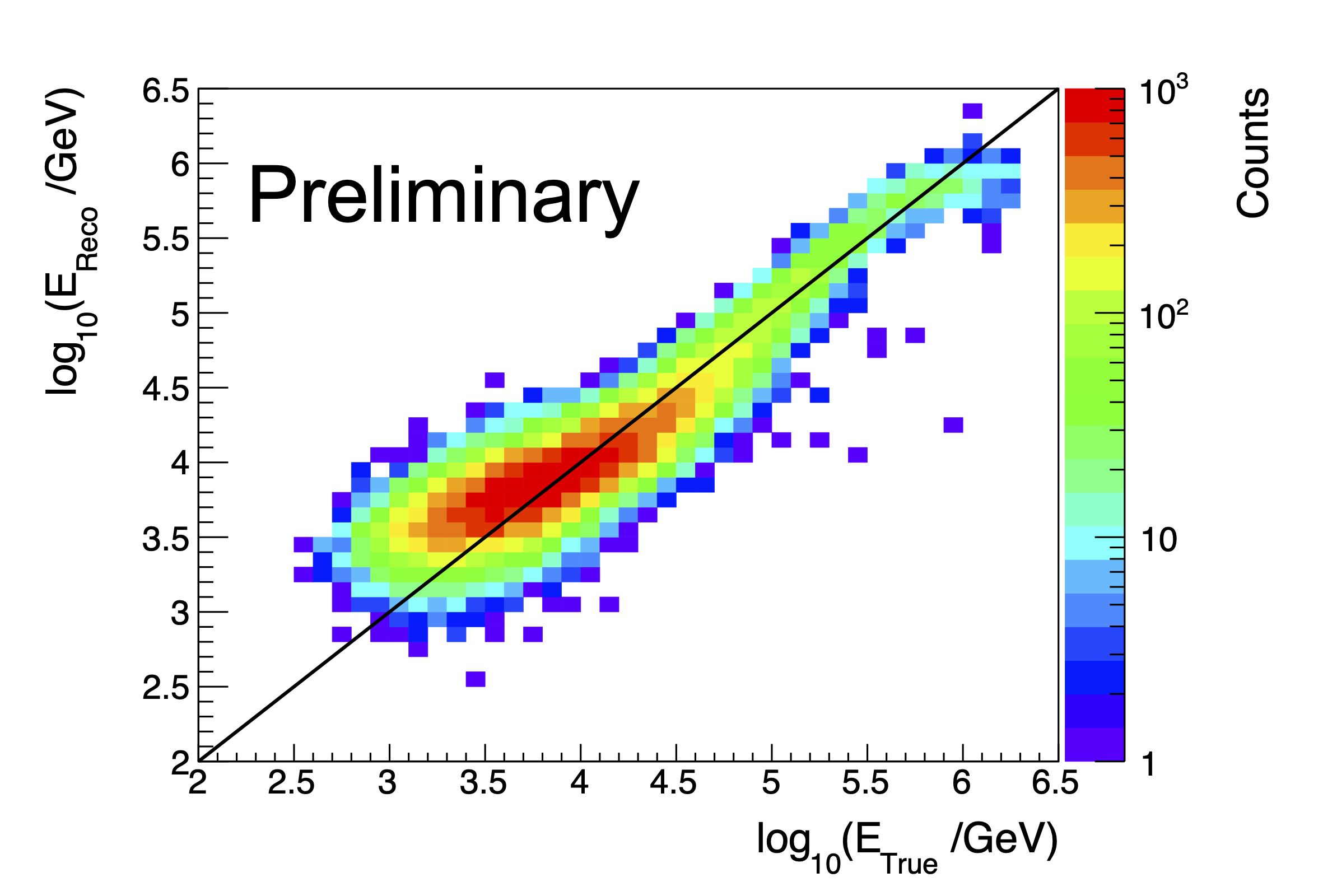}
         \caption{1st Neural Network set}
         \label{fig:nns1h2dsc}
     \end{subfigure}
     
     \medskip
     \begin{subfigure}{0.49\textwidth}
         \includegraphics[width=\textwidth]{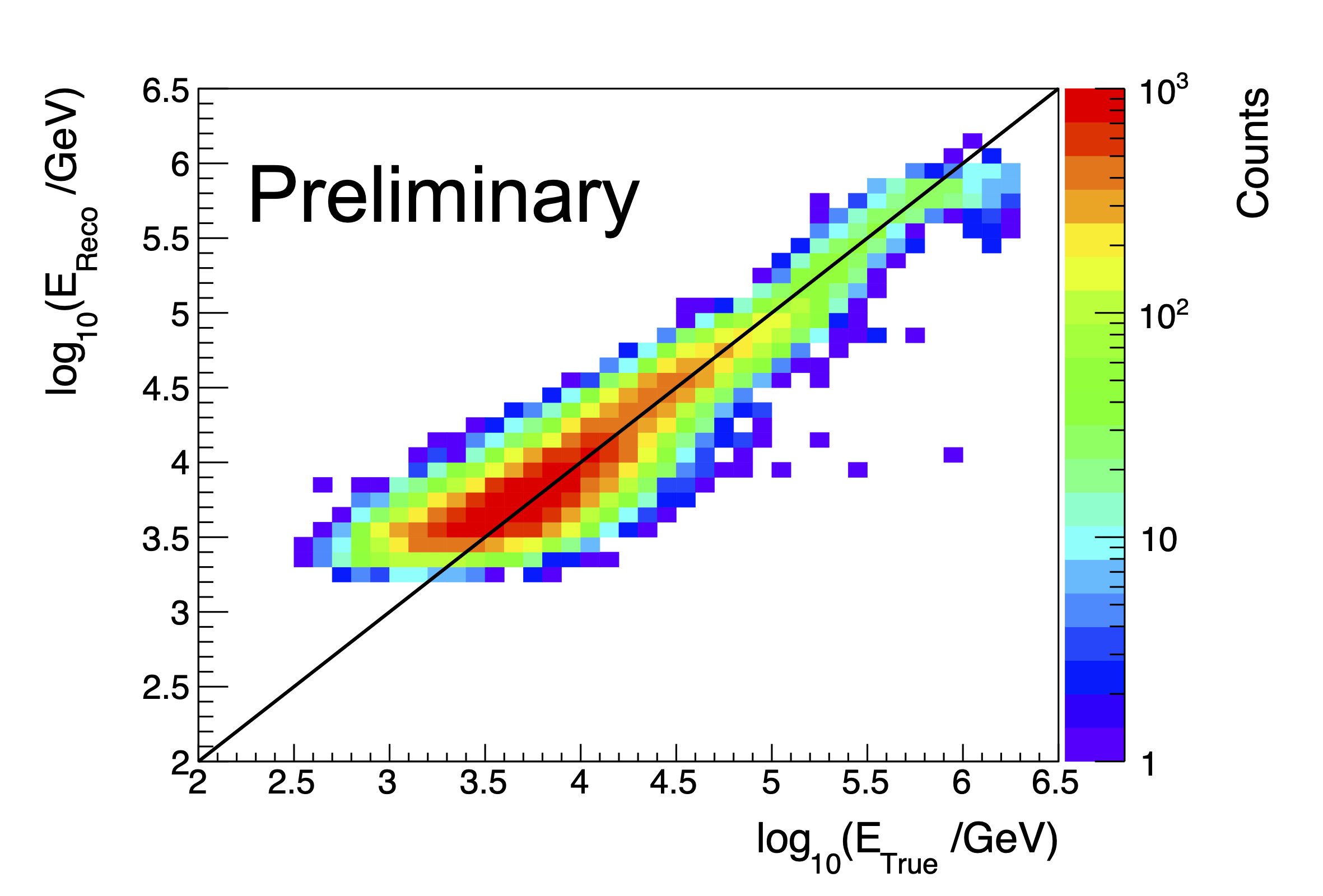
         }
         \caption{2nd Neural Network set}
         \label{fig:nns2h2dsc}
     \end{subfigure}
     \hfill
     \begin{subfigure}{0.49\textwidth}
         \includegraphics[width=\textwidth]{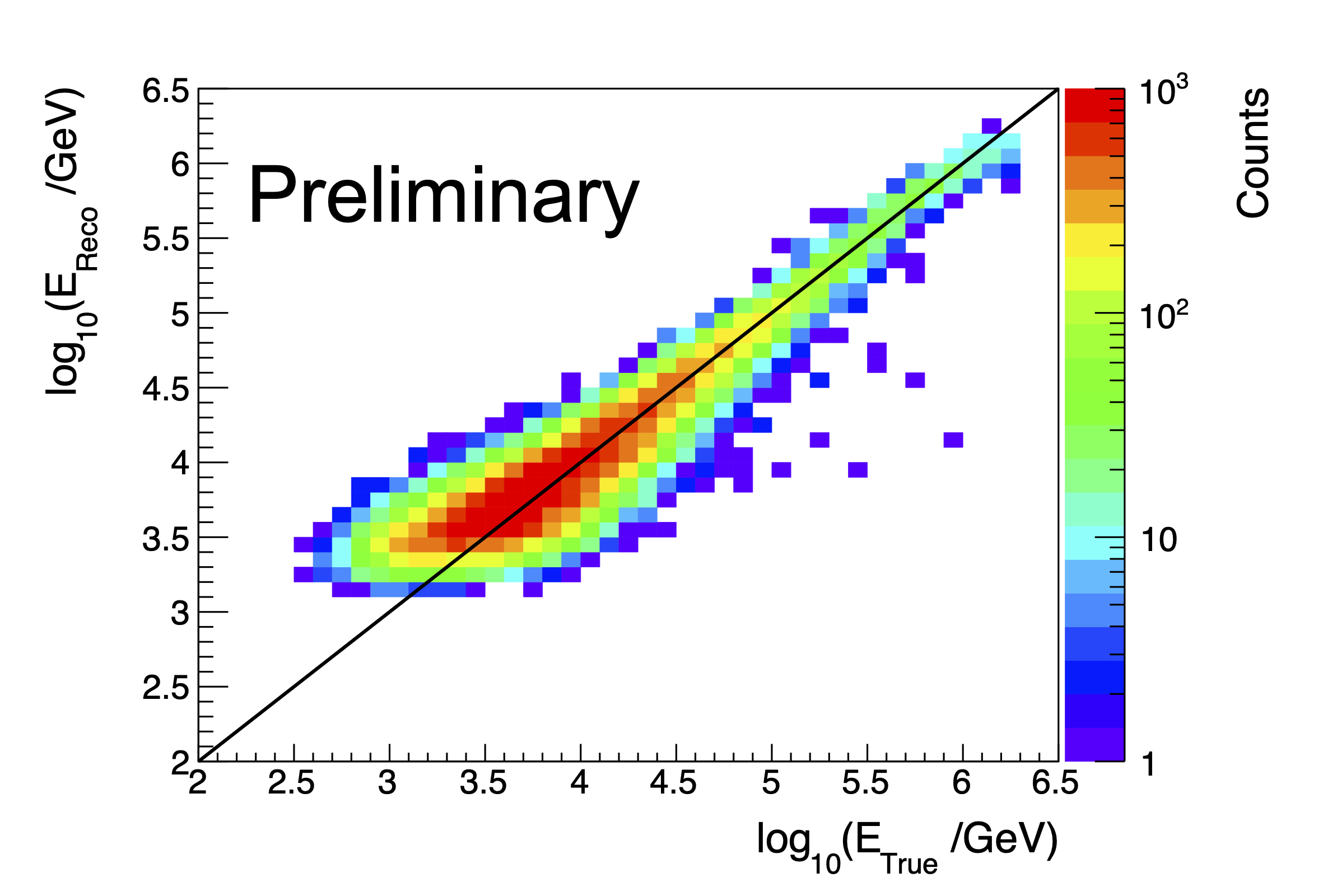}
         \caption{3rd Neural Network set}
         \label{fig:nns3h2dsc}
     \end{subfigure}
    
     \caption{A density heatmap is shown, illustrating the true energy versus the reconstructed energy using the Likelihood method (a), and the 1st, 2nd, and 3rd Neural Network sets (b, c, and d respectively) for testing proton-induced showers. The identity line is depicted by the black line.}
     \label{fig:modcom}
    
    \end{figure}

    \paragraph{}The optimal model is saved in a file with XML format after the training stage is completed. With this file, we can predict the energy for any event. To evaluate the model's performance, we use one-third of the total proton events as a test data set and compare the predictions. Figure \ref{fig:modcom} illustrates the distribution of reconstructed energy versus true energy for the vertical events (zenith angle $<17^{\circ}$) of the three model sets and the Likelihood. The best model is characterized by most events being closer to the identity line (solid dark line), indicating a more accurate prediction. From this results, we found that the third neural network set has the best performance (Figure \ref{fig:nns3h2dsc}). The Likelihood exhibits two undesired behaviors, both at high energies ($>100$~TeV). The first one is an underestimation of high-energy events, while in the second one is a loss of sensitivity at energies close to 1 PeV and above (resulting in a flat region in the top of Figure \ref{fig:lhh2dsc}).

\section{Testing stage}\label{sec:test}

\begin{figure}
         \begin{subfigure}{0.49\textwidth}
             \includegraphics[width=\textwidth]{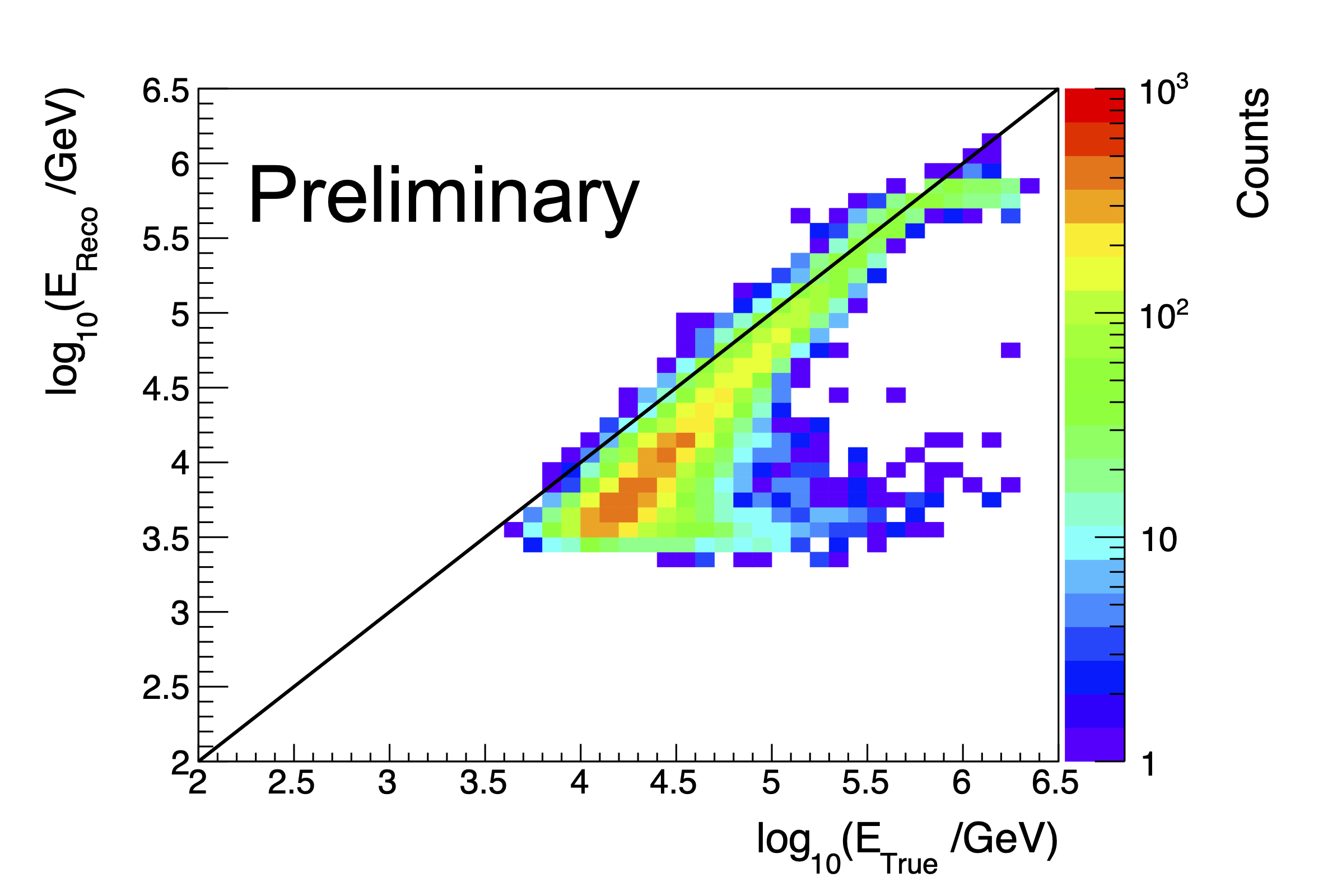}
             \caption{likelihood}
             \label{fig:lhh2dqcpi}
         \end{subfigure}
         \hfill
         \begin{subfigure}{0.49\textwidth}
             \includegraphics[width=\textwidth]{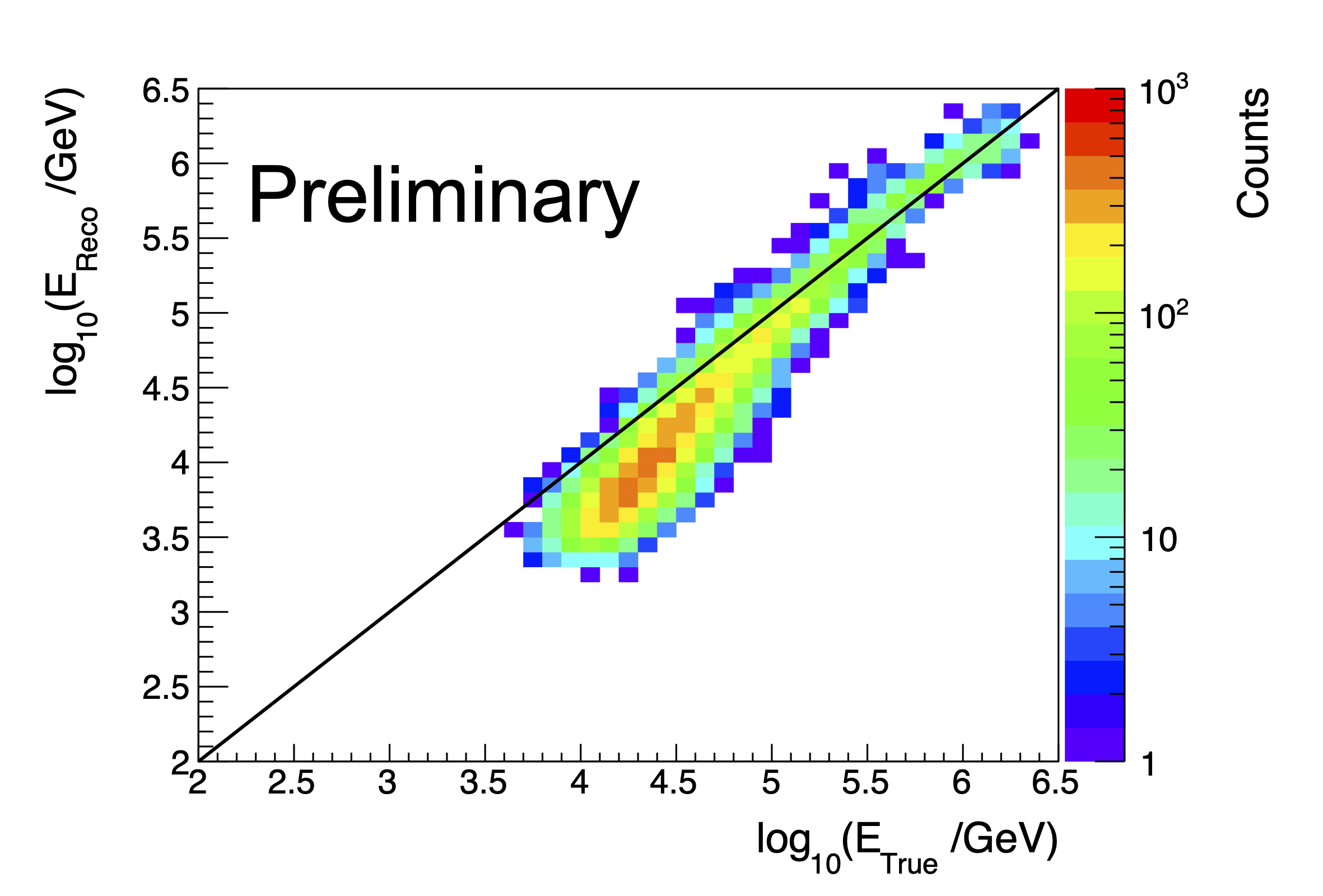}
             \caption{3rd Neural Network set}
             \label{fig:nns3h2dqcpi}
         \end{subfigure}
        
         \caption{A density heatmap is shown, illustrating the true energy versus the reconstructed energy using the Likelihood method (a) and the 3rd Neural Network set (b) for iron-induced showers. The identity line is depicted by the black line.}
         \label{fig:perh2d}
        
        \end{figure}

    \paragraph{}To assess the robustness of our neural network models for other cosmic ray nuclei in the detector, we conducted a test using the iron-induced shower (a heavier component). In figure \ref{fig:perh2d}, we compare the resulting performances for the Likelihood and the third neural network set. We observe that, in the case of iron nuclei, the Likelihood exhibits the same behavior as for the case of proton specie. However, the neural network set does not exhibit this behavior at high energies.
    
\begin{figure}
        \centering
         \begin{subfigure}{0.7\textwidth}
             \includegraphics[width=\textwidth]{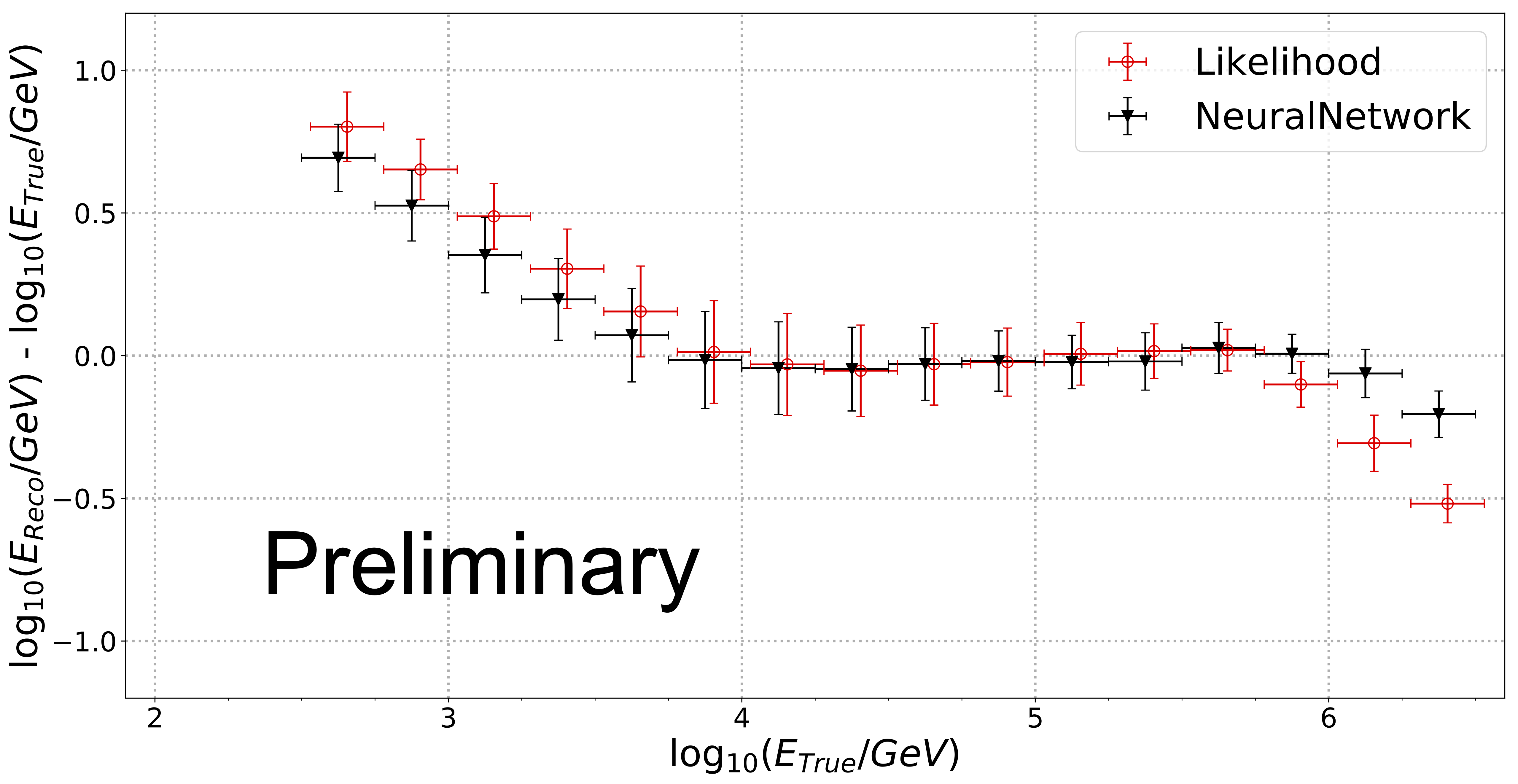}
             \caption{Proton-induced showers}
             \label{fig:proqcpp}
         \end{subfigure}
         
         \medskip
         \begin{subfigure}{0.7\textwidth}
             \includegraphics[width=\textwidth]{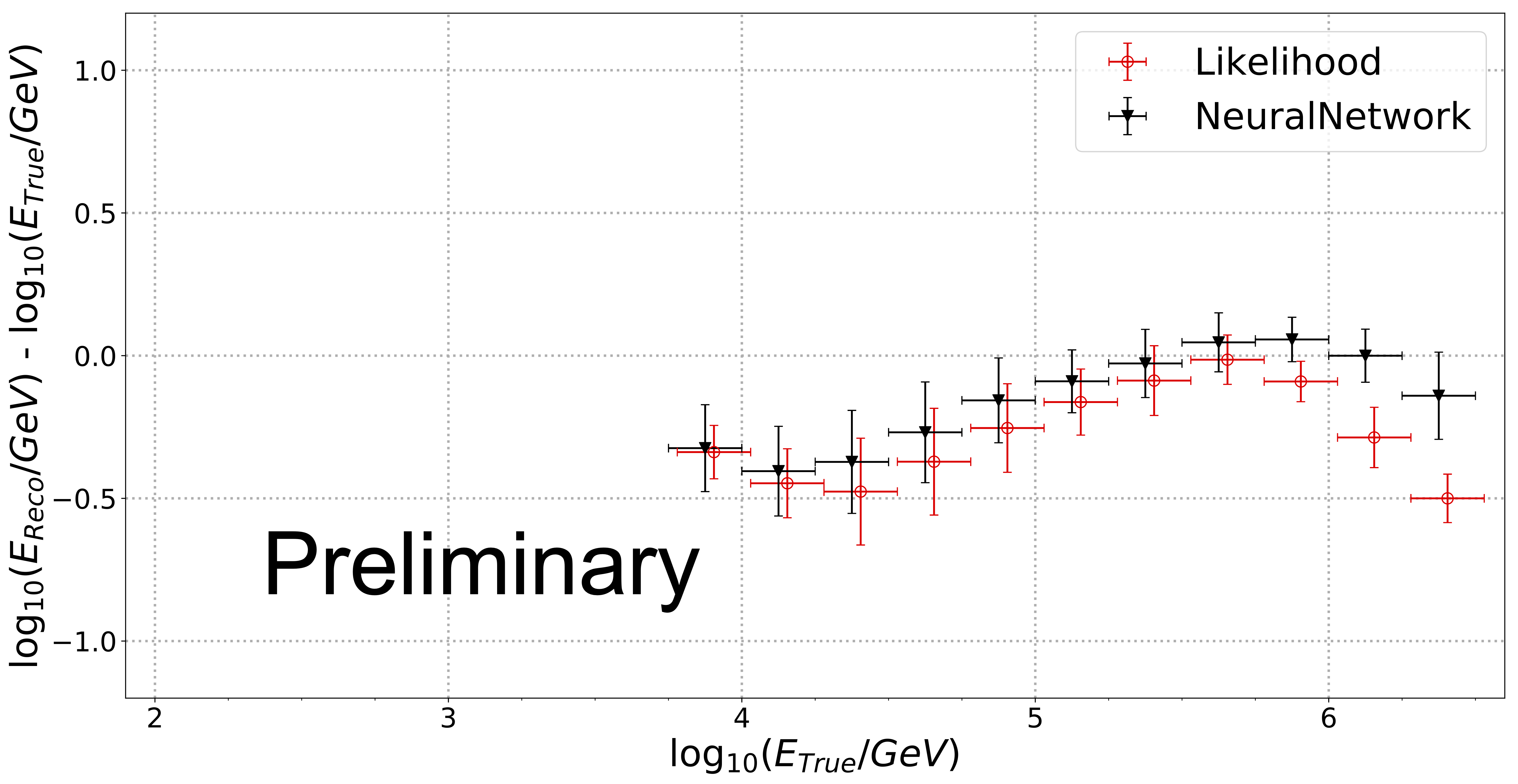}
             \caption{Iron-induced showers}
             \label{fig:proqcpi}
         \end{subfigure}
         \caption{A comparison is made between the bias associated with HAWC's official energy estimator Likelihood (red) and the 3rd Neural Network set (black) for two specie: (a) proton and (b) iron nuclei.}
         \label{fig:tprofile}
        
        \end{figure}

    \paragraph{}In order to quantify and evaluate the performance of the models, we calculate the bias, which is defined as the difference between the reconstructed and true energy values, both in logarithmic scale.
    \begin{equation}
        \text{bias}= \Delta \log_{10}(\text{E}) = \log_{10}\left(\text{E}_{\text{Reco}}/\text{GeV}\right) -~\log_{10}\left(\text{E}_{\text{True}}/\text{GeV}\right)
    \end{equation}
    We report the mean of the bias distribution for each quarter of a decade stating from $2.0$ ($100$~GeV). These results are presented in Figure~\ref{fig:tprofile}, considering the proton and iron-induced showers. The bias indicates the proximity of the predictions to the true energy. Between 10~TeV and 100~TeV, most events show a reconstructed energy value lower than the true one for both methods. Between 100~TeV and 1~PeV, both estimators demonstrate an excellent reconstruction, which is consistent with Figures \ref{fig:modcom} and \ref{fig:perh2d}, where the majority of events (indicated by the red zone) are close to the identity line. Finally, beyond 1 PeV, the neural network aligns the events with the identity line, as the bias approaches zero instead of having a significant offset, unlike the Likelihood estimator in the flat region.
    
\section{Discussion and Conclusions}\label{sec:conclu}

    \paragraph{}In this contribution, we reported the results for the energy estimator of cosmic-ray induced air showers with  three neural network sets trained using only proton specie. These sets share the same configuration parameters, such as architecture, number of epochs, and input variables, among characteristics. The difference lies in the number of networks trained within each set. The best model set is achieved when a network is specifically focused on reconstructing events within low, medium, and high fractional hit bins. This set exhibits significant improvements compared to the Likelihood model. In particular, the neural network model reduces the bias at energies close to 1 PeV and also reduce the size of the fluctuation at high energies. We obtain the same behavior when a heavier component, iron nuclei, is used.

    \paragraph{}Upon exploring the reconstruction of the iron event in the testing stage, it was concluded that it is advisable to train the model with multiple specie instead of just one. This is because there is an offset observed in estimated primary energy with respect to the true value, which is also observed with the Likelihood model. In general, the conclusion of the present study is that the third neural network set proves to be the superior model in comparison with the other ones explored in this work, showing notable improvement, particularly at high energies.

    \paragraph{}We will explore adding more variables that improve the reconstruction or training the model with a broader range of nuclei instead of relying solely on protons. This approach may help decrease bias in the reconstruction process. Another possibility is to train the models to focus on specific zenith angle bands or explore more sophisticated methods such as deep learning.

\begin{acknowledgments}
We acknowledge the support from: the US National Science Foundation (NSF); the US Department of Energy Office of High-Energy Physics; the Laboratory Directed Research and Development (LDRD) program of Los Alamos National Laboratory; Consejo Nacional de Ciencia y Tecnolog\'{i}a (CONACyT), M\'{e}xico, grants 271051, 232656, 260378, 179588, 254964, 258865, 243290, 132197, A1-S-46288, A1-S-22784, CF-2023-I-645, c\'{a}tedras 873, 1563, 341, 323, Red HAWC, M\'{e}xico; DGAPA-UNAM grants IG101323, IN111716-3, IN111419, IA102019, IN106521, IN110621, IN110521 , IN102223; VIEP-BUAP; PIFI 2012, 2013, PROFOCIE 2014, 2015; the University of Wisconsin Alumni Research Foundation; the Institute of Geophysics, Planetary Physics, and Signatures at Los Alamos National Laboratory; Polish Science Centre grant, DEC-2017/27/B/ST9/02272; Coordinaci\'{o}n de la Investigaci\'{o}n Cient\'{i}fica de la Universidad Michoacana; Royal Society - Newton Advanced Fellowship 180385; Generalitat Valenciana, grant CIDEGENT/2018/034; The Program Management Unit for Human Resources \& Institutional Development, Research and Innovation, NXPO (grant number B16F630069); Coordinaci\'{o}n General Acad\'{e}mica e Innovaci\'{o}n (CGAI-UdeG), PRODEP-SEP UDG-CA-499; Institute of Cosmic Ray Research (ICRR), University of Tokyo. H.F. acknowledges support by NASA under award number 80GSFC21M0002. We also acknowledge the significant contributions over many years of Stefan Westerhoff, Gaurang Yodh and Arnulfo Zepeda Dominguez, all deceased members of the HAWC collaboration. Thanks to Scott Delay, Luciano D\'{i}az and Eduardo Murrieta for technical support.
\end{acknowledgments}

\bibliographystyle{unsrt} 
	\bibliography{biblio}
\clearpage
\section*{Full Authors List: HAWC Collaboration}
\scriptsize
\noindent
%
\vskip2cm
\noindent
A. Albert$^{1}$,
R. Alfaro$^{2}$,
C. Alvarez$^{3}$,
A. Andrés$^{4}$,
J.C. Arteaga-Velázquez$^{5}$,
D. Avila Rojas$^{2}$,
H.A. Ayala Solares$^{6}$,
R. Babu$^{7}$,
E. Belmont-Moreno$^{2}$,
K.S. Caballero-Mora$^{3}$,
T. Capistrán$^{4}$,
S. Yun-Cárcamo$^{8}$,
A. Carramiñana$^{9}$,
F. Carreón$^{4}$,
U. Cotti$^{5}$,
J. Cotzomi$^{26}$,
S. Coutiño de León$^{10}$,
E. De la Fuente$^{11}$,
D. Depaoli$^{12}$,
C. de León$^{5}$,
R. Diaz Hernandez$^{9}$,
J.C. Díaz-Vélez$^{11}$,
B.L. Dingus$^{1}$,
M. Durocher$^{1}$,
M.A. DuVernois$^{10}$,
K. Engel$^{8}$,
C. Espinoza$^{2}$,
K.L. Fan$^{8}$,
K. Fang$^{10}$,
N.I. Fraija$^{4}$,
J.A. García-González$^{13}$,
F. Garfias$^{4}$,
H. Goksu$^{12}$,
M.M. González$^{4}$,
J.A. Goodman$^{8}$,
S. Groetsch$^{7}$,
J.P. Harding$^{1}$,
S. Hernandez$^{2}$,
I. Herzog$^{14}$,
J. Hinton$^{12}$,
D. Huang$^{7}$,
F. Hueyotl-Zahuantitla$^{3}$,
P. Hüntemeyer$^{7}$,
A. Iriarte$^{4}$,
V. Joshi$^{28}$,
S. Kaufmann$^{15}$,
D. Kieda$^{16}$,
A. Lara$^{17}$,
J. Lee$^{18}$,
W.H. Lee$^{4}$,
H. León Vargas$^{2}$,
J. Linnemann$^{14}$,
A.L. Longinotti$^{4}$,
G. Luis-Raya$^{15}$,
K. Malone$^{19}$,
J. Martínez-Castro$^{20}$,
J.A.J. Matthews$^{21}$,
P. Miranda-Romagnoli$^{22}$,
J. Montes$^{4}$,
J.A. Morales-Soto$^{5}$,
M. Mostafá$^{6}$,
L. Nellen$^{23}$,
M.U. Nisa$^{14}$,
R. Noriega-Papaqui$^{22}$,
L. Olivera-Nieto$^{12}$,
N. Omodei$^{24}$,
Y. Pérez Araujo$^{4}$,
E.G. Pérez-Pérez$^{15}$,
A. Pratts$^{2}$,
C.D. Rho$^{25}$,
D. Rosa-Gonzalez$^{9}$,
E. Ruiz-Velasco$^{12}$,
H. Salazar$^{26}$,
D. Salazar-Gallegos$^{14}$,
A. Sandoval$^{2}$,
M. Schneider$^{8}$,
G. Schwefer$^{12}$,
J. Serna-Franco$^{2}$,
A.J. Smith$^{8}$,
Y. Son$^{18}$,
R.W. Springer$^{16}$,
O.~Tibolla$^{15}$,
K. Tollefson$^{14}$,
I. Torres$^{9}$,
R. Torres-Escobedo$^{27}$,
R. Turner$^{7}$,
F. Ureña-Mena$^{9}$,
E. Varela$^{26}$,
L. Villaseñor$^{26}$,
X. Wang$^{7}$,
I.J. Watson$^{18}$,
F. Werner$^{12}$,
K.~Whitaker$^{6}$,
E. Willox$^{8}$,
H. Wu$^{10}$,
H. Zhou$^{27}$

\vskip2cm
\noindent
$^{1}$Physics Division, Los Alamos National Laboratory, Los Alamos, NM, USA,
$^{2}$Instituto de Física, Universidad Nacional Autónoma de México, Ciudad de México, México,
$^{3}$Universidad Autónoma de Chiapas, Tuxtla Gutiérrez, Chiapas, México,
$^{4}$Instituto de Astronomía, Universidad Nacional Autónoma de México, Ciudad de México, México,
$^{5}$Instituto de Física y Matemáticas, Universidad Michoacana de San Nicolás de Hidalgo, Morelia, Michoacán, México,
$^{6}$Department of Physics, Pennsylvania State University, University Park, PA, USA,
$^{7}$Department of Physics, Michigan Technological University, Houghton, MI, USA,
$^{8}$Department of Physics, University of Maryland, College Park, MD, USA,
$^{9}$Instituto Nacional de Astrofísica, Óptica y Electrónica, Tonantzintla, Puebla, México,
$^{10}$Department of Physics, University of Wisconsin-Madison, Madison, WI, USA,
$^{11}$CUCEI, CUCEA, Universidad de Guadalajara, Guadalajara, Jalisco, México,
$^{12}$Max-Planck Institute for Nuclear Physics, Heidelberg, Germany,
$^{13}$Tecnologico de Monterrey, Escuela de Ingeniería y Ciencias, Ave. Eugenio Garza Sada 2501, Monterrey, N.L., 64849, México,
$^{14}$Department of Physics and Astronomy, Michigan State University, East Lansing, MI, USA,
$^{15}$Universidad Politécnica de Pachuca, Pachuca, Hgo, México,
$^{16}$Department of Physics and Astronomy, University of Utah, Salt Lake City, UT, USA,
$^{17}$Instituto de Geofísica, Universidad Nacional Autónoma de México, Ciudad de México, México,
$^{18}$University of Seoul, Seoul, Rep. of Korea,
$^{19}$Space Science and Applications Group, Los Alamos National Laboratory, Los Alamos, NM USA
$^{20}$Centro de Investigación en Computación, Instituto Politécnico Nacional, Ciudad de México, México,
$^{21}$Department of Physics and Astronomy, University of New Mexico, Albuquerque, NM, USA,
$^{22}$Universidad Autónoma del Estado de Hidalgo, Pachuca, Hgo., México,
$^{23}$Instituto de Ciencias Nucleares, Universidad Nacional Autónoma de México, Ciudad de México, México,
$^{24}$Stanford University, Stanford, CA, USA,
$^{25}$Department of Physics, Sungkyunkwan University, Suwon, South Korea,
$^{26}$Facultad de Ciencias Físico Matemáticas, Benemérita Universidad Autónoma de Puebla, Puebla, México,
$^{27}$Tsung-Dao Lee Institute and School of Physics and Astronomy, Shanghai Jiao Tong University, Shanghai, China,
$^{28}$Erlangen Centre for Astroparticle Physics, Friedrich Alexander Universität, Erlangen, BY, Germany


\end{document}